# Nonlinear exciton-Mie coupling in transition metal dichalcogenides nanoresonators

*Anna A. Popkova   Ilya M. Antropov    Gleb I. Tselikov         Georgy A. Ermolaev   Igor Ozerov    Roman V. Kirtaev         Sergey M. Novikov     Andrey B. Evlyukhin  Aleksey V. Arsenin  Vladimir O. Bessonov       Valentyn S. Volkov        Andrey A. Fedyanin**

A. A. Popkova, I. M. Antropov, V. O. Bessonov
Faculty of Physics, Lomonosov Moscow State University, Moscow, 119991, Russia
G. I. Tselikov, G. A. Ermolaev, R. V. Kirtaev, S. M. Novikov, A. V. Arsenin, V. S. Volkov
Center for Photonics and 2D Materials, Moscow Institute of Physics and Technology, 9 Institutsky Lane, Dolgoprudny, 141700, Russian Federation
I. Ozerov
Aix-Marseille Univ, CNRS, CINAM, 13288 Marseille, France
A. B. Evlyukhin
Institute of Quantum Optics, Liebniz University Hannover, 30167 Hannover, Germany
A. V. Arsenin, V. S. Volkov
GrapheneTek LLC, Moscow 109004, Russia
A. A. Fedyanin
Faculty of Physics, Lomonosov Moscow State University, Moscow, 119991, Russia Email Address: fedyanin@nanolab.phys.msu.ru

Thanks to a high refractive index, giant optical anisotropy, and pronounced excitonic response, bulk transition metal dichalcogenides (TMDCs) have recently been discovered to be an ideal foundation for post-silicon photonics. The inversion symmetry of bulk TMDCs, on the other hand, prevents their use in nonlinear-optical processes such as second-harmonic generation (SHG). To overcome this obstacle and broaden the application scope of TMDCs, we engineered $MoS_2$ nanodisks to couple Mie resonances with C-excitons. As a result, their alliance produced 23-fold enhancement of SHG intensity with respect to the resonant SHG from a high-quality exfoliated $MoS_2$ monolayer under C-exciton excitation. Furthermore, SHG demonstrates a strongly anisotropic response typical of a $MoS_2$ monolayer due to the single-crystal structure of the fabricated nanodisks, providing a polarization degree of freedom to manipulate SHG. Hence, our results significantly improve the potential of bulk TMDCs enabling an avenue for next-generation nonlinear photonics.

## 1  Introduction

Owing to the enormous potential in post-silicon on-chip technology [1, 2], bulk transition metal dichalcogenides (TMDCs) have emerged from the shadows of monolayer counterparts in recent years. Ultrafast photodetection [3, 4], exciton-polariton transport [1, 5, 6], strong coupling [7, 8], Zenneck surface waves [9], tunable birefringence [10], anapole modes [11, 12], and ultrasensitive sensors [13, 14] are some of the most well-known examples. Superior bulk TMDC characteristics, such as high refractive index [15], excitonic light-matter interaction [16], and giant optical anisotropy [1], lie at the core of these advancements. However, unlike monolayers [17], inversion symmetry in bulk TMDC leads to negligible quadratic nonlinear optical response, limiting their applications in high harmonic generation [18], switching [19], high-resolution imaging [20], and terahertz generation [21]. As a result, providing a sophisticated optical design for the use of TMDCs in nonlinear processes is critical. Recent studies suggest using different TMDC morphologies, such as nanomesh [22], tubular [23], and nanoporous [24] networks.

Despite the excellent findings of these studies, these nanostructures are challenging to incorporate into optical circuits [25]. As a result, a new technological approach for TMDC nonlinear integrated nanophotonics is in high demand.

All-dielectric resonant metaphotonics [26], fortunately, gives a hint to this quest. Manipulation of strong electromagnetic Mie-type resonances in high refractive index nanostructures, for example, considerably amplifies linear [11] and nonlinear [27, 28, 29, 30, 31] optical responses of materials. Another advantage of TMDCs is their excitonic nature of dielectric response [16, 32], which greatly enhances light-matter interaction [33]. Specifically, this leads to an increase in harmonic generation due to resonances of nonlinear susceptibilities in the spectral vicinity of excitons both in the monolayer [34, 35] and bulk [36] of TMDCs. As a result, combining Mie-type resonances with TMDC excitons could result in significant harmonic generation. Indeed, our findings show that an innovative exciton-Mie coupling regime greatly boosts second-harmonic generation (SHG) in TMDC nanoresonators.

In this work, we designed and fabricated $MoS_2$ nanodisks, which support Mie-type resonances at the fundamental wavelength of 900 nm along with the C-exciton resonance at the second-harmonic (SH) wavelength of 450 nm. The proposed system demonstrates the SHG enhancement by more than an order of magnitude in comparison with the maximum achievable intensity of SHG in a $MoS_2$ monolayer under the C-exciton resonance. Interestingly, SHG exhibits highly anisotropic response, which tentatively originates from hexagonal crystalline structure of fabricated nanodisks. This anisotropic behavior provides an extra degree of freedom to control second-harmonic intensity [37]. Our exciton-Mie approach yields an indispensable route for nonlinear optics in TMDC next-generation photonics.

## 2 Experimental samples

A sketch of the system under study is shown in **Figure 1(a)**. The proposed idea is to excite a Mieresonance mode in a $MoS_2$ nanodisk at the fundamental frequency corresponding to half of the $MoS_2$ Cexciton frequency. In this case, the nonlinear exciton-Mie coupling can be understood from the expression for the SH field: $E_{2\omega} = \chi^{(2)}(2\omega)L^2(\omega)E_\omega^2$, where $\chi^{(2)}$ is the $MoS_2$ quadratic susceptibility depending on the light frequency and $L$ is a local field factor indicating the amplification of the pump field inside the material compared to the incident field $E_\omega$. One can dramatically enhance SHG at specific frequency by combining the resonance of the quadratic susceptibility with the Mie-type resonance of the local field factor tailored by the parameters of the nanodisk. We chose the C-exciton resonance, in the spectral vicinity of which the highest value of the quadratic susceptibility has been observed so far for a $MoS_2$ monolayer [34], as well as the significant enhancement of SHG has been shown in a centrosymmetric bilayer [38, 39] and thin films of $MoS_2$ [36]. The disk shape of nanoresonators allows us to flexibly design resonant modes due to two degrees of freedom: height and diameter. The very high absorption in the spectral region of the C-exciton [1] prevents the excitation of Mie-modes in the nanodisk at the SH frequency and leads to efficient SHG only from the disk's upper layer with the thickness of about 10 nm. Thus, the field of the Mie-mode generating the SH radiation should be localized near the top of the nanodisk. Using numerical simulation by finite difference in time-domain (FDTD) method taking into account the $MoS_2$ optical anisotropy [1], we determined that the excitation of the magnetic dipole (MD) mode is accompanied by an increase in the in-plane electric field components at the upper surface of the nanodisk. The disks height is determined by the height of the $MoS_2$ flake being 110 nm, thus the spectral position of the MD resonances was adjusted by changing the disk diameter. For the disk diameter of 550 nm, MD mode is excited at the wavelength of 900 nm corresponding to twice the C-exciton wavelength. As a reference sample, we chose nanodisks with the diameter of 300 nm with the MD resonance observed at the pump wavelength of 800 nm.

$MoS_2$ nanodisks were fabricated from single-crystal $MoS_2$ flakes placed on a silicon wafer covered by 290 nm-thick thermally grown silicon oxide layer. As a fabrication method, we used a well-controlled combination of electron beam lithography and etching with a negative resist as an etching mask allowing us to completely remove remaining $MoS_2$ around the disks.

The optical dark-field image of the sample is shown in Figure 1(b). Light lines are the boundaries of the etched flake appeared since the silicon oxide layer outside the flake was over-etched simultaneously with $MoS_2$. Each bright spot inside the boundaries corresponds to a $MoS_2$ nanodisk. Figure 1(c) shows the AFM image of the sample. A part of the unwashed resist is seen atop of the disk, which is challenging to remove due to the weak attachment of the $MoS_2$ flake to the substrate. Using AFM and SEM measurements, we revealed that nanodisks have slightly inclined vertical walls and a shape close to hexagonal, caused by preferential etching of structures along the armchair crystalline direction [40]. The particle size was determined as the diameter of the inscribed circle at the top of the nanodisk and found to be very close to values determined from the calculation (see Supporting Information, Section I for details). Nevertheless, the non-circular shape and unwashed resist does not affect the spectral position of the MD resonance and the field distribution inside the nanodisk (see Supporting Information, Section III). The Raman spectra demonstrate the bulk crystallinity of the fabricated nanodisks (see Supplementary Information, Section I). The presence of excitons in $MoS_2$ nanodisks was confirmed by measuring the reflection spectra of the samples, in which excitons manifest themselves as resonances at 460-, 615- and 680-nm wavelengths [41] (see Figure S2 from Supplementary Information).

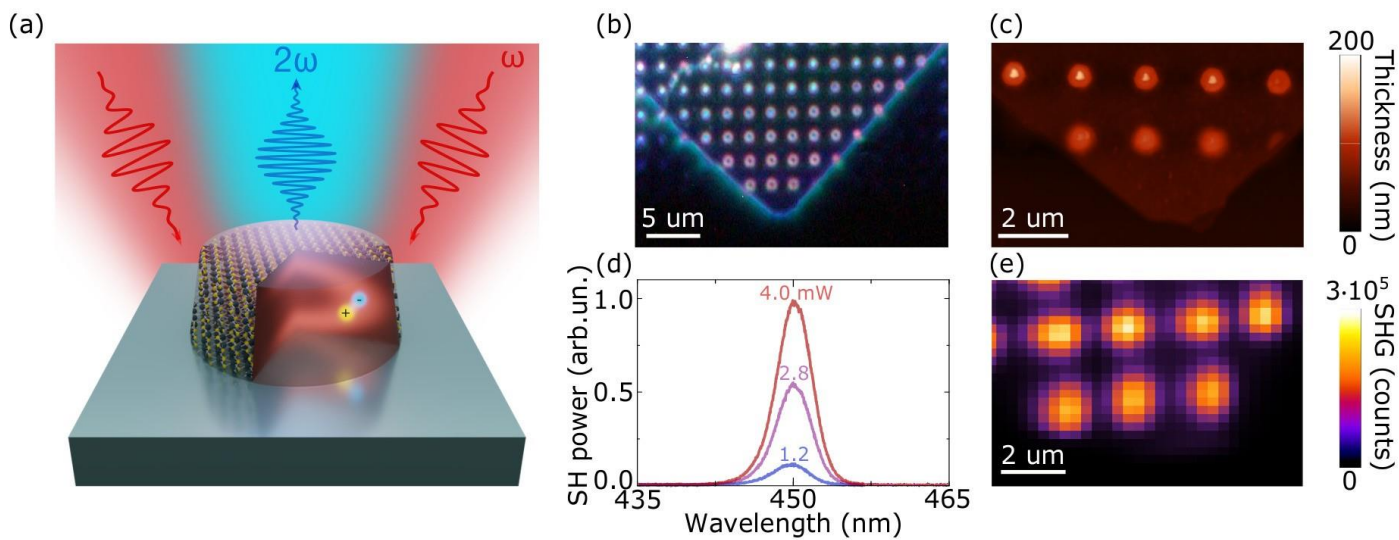

Figure 1: (a) Schematic of a $MoS_2$ nanoresonator and SHG experiment. Pump radiation excites a Mie mode at half the frequency of the exciton resonance of the quadratic susceptibility, thus boosting SHG. (b) Optical dark-field microscopy image of the sample with 550-nm nanodisks. (c) AFM image of the studied area. (d) Spectra of the SH signal from the single 550-nm nanodisk for various pump powers. (e) SH intensity map of the studied area with 550-nm $MoS_2$ nanodisks, measured at a pump wavelength of 900 nm.

## 3 Results and discussion

The nonlinear-optical signal reflected from the sample was measured using a home-built multiphoton microscope based on a femtosecond laser tunable in the spectral range from 680 nm to 1080 nm (see

Supplementary Information, Section II for details) [42]. Figure 1(d) shows spectra of SH intensity generated by a single 550-nm nanodisk measured at various powers of the pump radiation at the fundamental wavelength of 900 nm. The spectra have a peak centered at 450 nm with a width equal to half the width of the pump pulse spectrum [43] and a zero background indicating the absence of multiphoton luminescence and other nonlinear processes in the detected spectral region (see SH spectrum in the wider spectral range in Figure S5). Figure 1(e) depicts the SH intensity map of the sample area shown in Figure 1(c) measured by scanning with the 2-$\mu$m diameter pump beam in increments of 0.3 $\mu$m. The SH intensity maxima correspond exactly to the positions of the MoS$_2$ disks and exceed the SH signal from the substrate by three orders of magnitude.

Nonlinear-optical methods such as SHG are well-suited to find out the symmetry and orientation of crystals. We measured the dependence of the SH intensity on the azimuthal angle of the sample rotation to check whether the crystal structure of MoS$_2$ was affected during the etching process. The SHG anisotropy dependences for a single MoS$_2$ nanodisk are shown in **Figure 2(a)**. Blue and black dots correspond to the parallel and perpendicular orientations of the SH polarization relative to the polarization direction of the pump radiation. Each point of the graph corresponds to the signal from the same disk, measured by scanning the sample for various azimuthal angles. The typical SH scans for several angles are depicted in Figure 2(b). The SH intensity from the nanoantenna exhibits strongly varying, 6-fold symmetric response as a function of azimuthal angle for both polarizations, that corresponds to the D$_{3h}$ symmetry group of the MoS$_2$ monolayer [41]. The SH intensity at the minimum is at least two orders of magnitude less than the maximum value, indicating the absence of a significant isotropic contribution. This suggests that the resonators save the crystal structure of MoS$_2$ during fabrication.

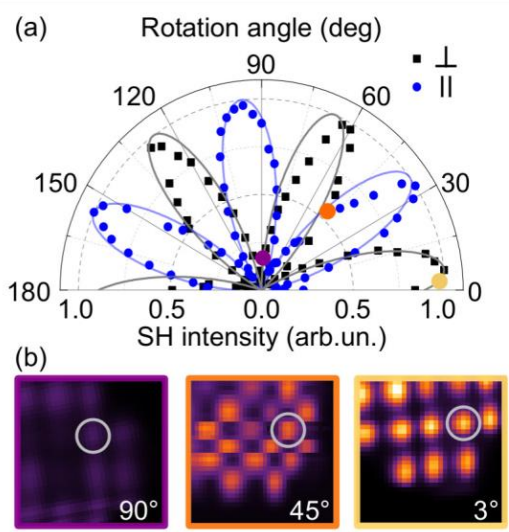

Figure 2: (a) Polar plot of the SH intensity measured for the individual nanodisk as a function of the sample azimuthal angle. The polarization of the SHG radiation is parallel (blue dots) and perpendicular (black dots) to the polarization of the pump radiation. Solid curves correspond to the approximation of the experimental data by the $sin^2(3\varphi)$ function, where $\varphi$ is the sample rotation angle. (b) SH intensity maps of the same sample area for various azimuthal angles. The gray circle marks the nanodisk, for which the anisotropic dependence is shown in panel (a). Three colored circles in panel (a) indicate the intensity of SHG from marked nanodisk for the cases shown in panel (b).

To measure the SH spectra from single MoS$_2$ nanodisks, we illuminated the samples with wide pump beam 15 $\mu$m in diameter and detected a diffraction-limited image of the disk array at the SH wavelength (see Supporting Information, Section II for details). The SH images obtained at the resonant fundamental wavelengths are shown in **Figure 3(a)** confirming that SH is generated solely in nanodisks. Figure 3(b) displays the spectral dependence of the power of SH radiation generated by the disks marked with colored circles in Figure 3(a). Black dots show the SH spectrum generated in a MoS₂ monolayer using the 2 µm pump beam with the same fluence as in measurements with nanodisks. For the 550-nm disk, the SH spectrum has the narrow resonance near 900-nm fundamental wavelength. A similar spectral behavior is observed for the MoS₂ monolayer with the SH power being approximately three times smaller. The resonance of SHG from the monolayer is associated with the quadratic susceptibility resonance in the vicinity of the C-exciton [34]. For the disk with a diameter of 300 nm, SH spectrum has two resonances. The spectral position of the long-wavelength resonance corresponds to the C-exciton susceptibility resonance, however, the maximum SH gain is observed at the 800-nm pump wavelength. We measured the linear scattering spectra (Figure 3(c)) of the same nanodisks to compare the resonances observed in the linear and nonlinear response. For the larger disk, the scattering resonance is revealed near the 900-nm pump wavelength. The scattering spectrum of the smaller disk has the peak near the 800-nm wavelength. The results of FDTD simulation shown by curve in Figure 3(c) demonstrate good agreement with experimental data and confirm that the observed scattering resonances are associated with the excitation of MD modes in the nanodisks.

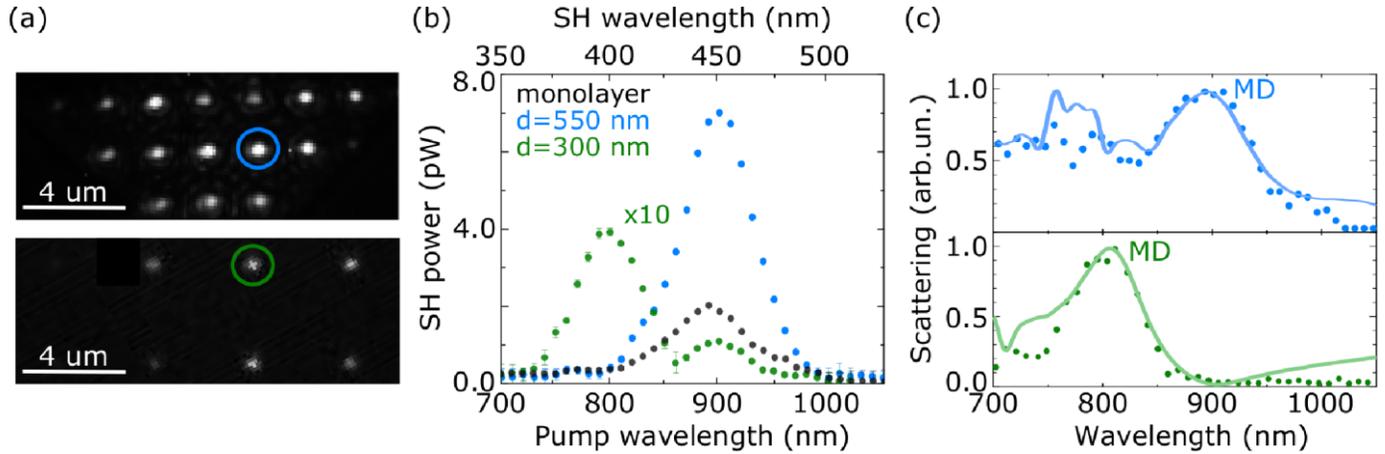

Figure 3: (a) SH images of 550-nm (top) and 300-nm (bottom) nanodisks, obtained at 900-nm and 800-nm pump wavelengths, respectively. Color circles indicate the nanodisks, the SHG and scattering spectra of which are shown in panels (b) and (c). (b) SHG spectra of MoS₂ nanodisks (color dots) and monolayer (black dots) measured at 16 GW/cm² peak intensity corresponding to the 5 mW pump power for 2-µm focused beam. (c) Measured (dots) and calculated (curves) scattering spectra of the MoS₂ nanodisks with a diameter of 550 nm (top) and 300 nm (bottom).

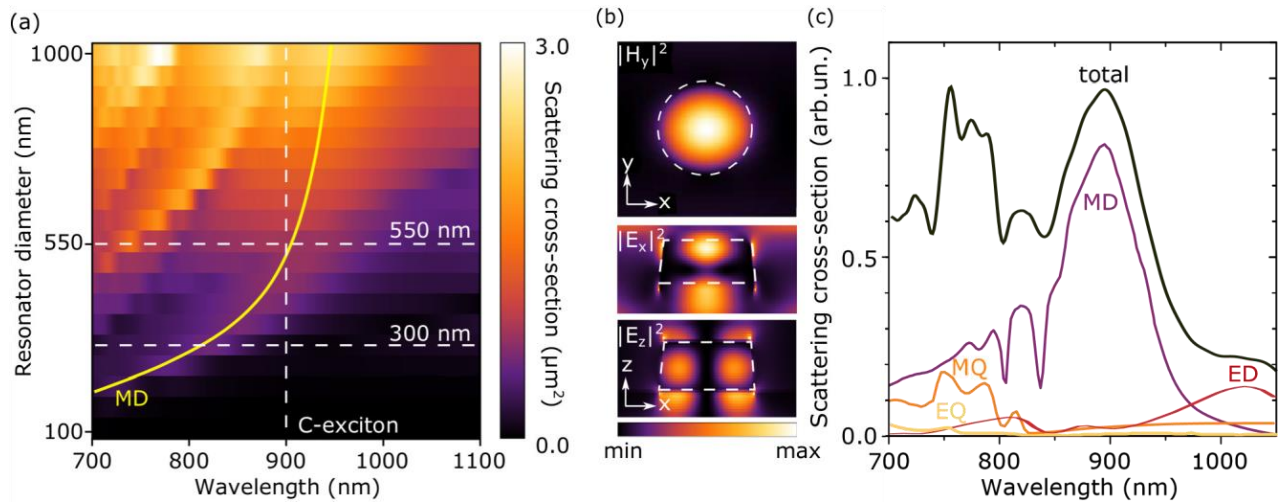

Figure 4: (a) Spectra of scattering cross-section calculated for various disk diameters. Yellow curve shows the position of the MD resonance. The dashed lines indicate the sizes of the experimental nanodisks and doubled C-exciton wavelength. (b) Electric and magnetic fields distributions in the 550-nm nanodisk at the 900-nm pump wavelength. Dashed curve indicates the disk boundaries. (c) Result of numerical multipole decomposition of scattering cross-section for the 550-nm nanodisk.

We performed numerical simulation for MoS$_2$ disks of various diameters to trace the spectral shift in the positions of Mie resonances. The calculated scattering cross-section is shown in **Figure 4(a)**. Analysis of electric and magnetic field distributions together with the results of numerical multipole decomposition [44] allow one to determine the spectral positions of individual Mie resonances. The yellow curve in Figure 4(a) depicts the dependence of the spectral position of MD resonance on the disk diameter. The field distribution for the 550-nm disk at the 900-nm wavelength is shown in Figure 4(b) and is typical for the MD mode in nanodisks. A similar fields distribution is observed for the 300-nm disk near the 800-nm wavelength (see Figure S8 in Supporting Information). The enhancement of the in-plane field component at the disk surface leads to efficient SHG in the near-surface MoS$_2$ layers and its emission into free space, while SH generated in the disk volume is absorbed by the material. An example of numerical multipole decomposition for the 550-nm nanodisk shown in Figure 4(c) demonstrates the sole contribution of MD resonance to the optical response in the vicinity of the doubled C-exciton wavelength. Note, that in the short-wavelength region, the sum of the contributions of the calculated multipoles is less than the total scattering cross section. This is explained by the contributions to scattering from the Si/SiO$_2$ substrate, higher order multipoles (octupoles, etc.), as well as possible Fabry–Pérot modes, which are not taken into account in the calculations. Nevertheless, a comparison of the numerical decomposition with the more rigorous analytical multipole decomposition performed by the discrete dipoles method [45, 46] for a MoS$_2$ nanodisk in vacuum shows good agreement between the spectral positions of the Mie resonances (see Supplementary Information, Section III).

The measured SH signal from monolayer allows us to estimate the second-order nonlinear optical susceptibility of MoS$_2$ at C-exciton resonance. Using the method described in Ref.[47], we obtain $\chi^{(2)}$=220 pm/V (see Supporting Information, Section IV) that is in a good agreement with previously reported one [34]. This value can be slightly underestimated due to the complex structure of the substrate (see Supporting Information, Section IV for details). Using $MoS_2$ monolayer as a reference, we can estimate the gain in the nonlinear response of nanodisks. To do this correctly, we must take into account the area of the SHG sources and obtain the SH intensity for each sample. For a disk, the SH source area can be estimated as the area of the disk's upper face, measured using scanning electron microscopy or AFM. In√ the case of monolayer, the SH source area is the pump spot area divided by 2. The spectral dependences of the SH intensity are shown in Figure5, where the maximum value of intensity for the $MoS_2$ monolayer is taken as the unity. The SH intensity gain for the 300-nm disk is approximately 5 when the MD mode is excited. The enhancement of the intensity at the wavelength of the C-exciton coincides in amplitude with the monolayer one. For double-resonant sample, the maximum SHG gain of 23 is observed at the 900-nm wavelength that corresponds to the overlap of MD resonance and C-exciton. In this case, the intensity is $10^3$ times greater than that generated in $MoS_2$ monolayer away from C-exciton resonance. The experimental conversion coefficient is $5\times10^{-7}\,W^{-1}$. Taking into account that only 25% of the pump power hits the nanodisk surface, the actual SH efficiency can be estimated as $2\times10^{-6}\,W^{-1}$. The observed values of the gain and conversion coefficients are comparable with the ones obtained using optomechanical amplification of the SH signal [48], as well as the combination of a $MoS_2$ monolayer with a micro-cavity [49] and bound state in the continuum [50] systems. Also, the demonstrated efficiency is two orders of magnitude larger than the SH conversion efficiency for a $MoS_2$ metasurface with Mie resonances [51] in the absence of absorption at the SH frequency.

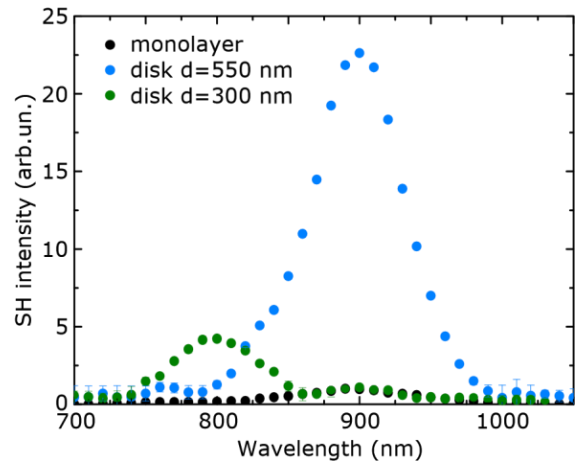

Figure 5: SH intensity spectra for the 550-nm (blue) and 300-nm (green) $MoS_2$ nanodisks and $MoS_2$ monolayer (black). The maximum intensity of SHG from monolayer at C-exciton resonance is taken as unity.

## Conclusion

In conclusion, the high refractive index of bulk TMDCs, along with the presence of excitons at room temperature, allows creating a nanoresonator with a significantly enhanced non-linear optical response. Here, thanks to flexible adjusting of Mie-resonance modes, we have couple the exciton resonance of the $MoS_2$ quadratic susceptibility with the MD resonance through a nonlinear optical process of SHG. By etching a single-crystal $MoS_2$ flake, nanodisks were fabricated to support the MD mode at a wavelength of 900 nm corresponding to the doubled C-exciton wavelength. As the result, the intensity of SH generated by the nanodisks exceeded the intensity of SHG from $MoS_2$ monolayer at the C-exciton resonance by a factor of 23. Single-crystal nanodisks exhibit 6-fold anisotropic SHG, similar to the monolayer case, which is an additional knob in controlling the SH radiation. A variety of subwavelength geometrical modes and their interplay in nanoresonators [27, 52, 53] can be used to further enhance the exciton-mediated nonlinear response. In turn, the electrical control of the exciton-enhanced quadratic susceptibility [38, 39, 19] paves the way towards a new tunable nonlinear nanophotonic devices based on TMDCs.

**Supporting Information**

Supporting Information is available from the Wiley Online Library or from the author.

**Acknowledgements**

The work was performed under partial financial support of the Russian Ministry of Education and Science (Grant No. 14.W03.31.0008), the Russian Science Foundation (grant No. 21-19-00675, sample fabrication and characterization, grant No. 20-12-00371, nonlinear-optical measurements, No. 20-12-00343, multipole decompositions), and Russian Foundation for Basic Research (grant No.21-52-12036, numerical simulations).

# Methods

### Sample fabrication

Arrays of $MoS_2$ nanoresonators were fabricated by electron beam lithography (EBL) followed by reactive ion etching (RIE). We used the substrates cleaved from silicon (100) wafers covered by a 290-nm-thick thermally grown silicon oxide. The substrates were pre-patterned by optical lithography in order to facilitate the identification of areas for the following $MoS_2$ deposition. Firstly, the substrates were cleaned in sequential ultrasonic bathes of acetone and isopropanol, rinsed in deionized water and dried under clean nitrogen flow. Then the substrates were exposed to the oxygen radical plasma treatment in a barrel reactor (Nanoplas, France). $MoS_2$ flakes were deposited on substrates using mechanical exfoliation from highly oriented synthetic 2H-phase $MoS_2$ crystal (2D Semiconductors Inc., USA). The flakes to pattern were chosen under an optical microscope and their thicknesses were measured by a contact profilometer (DektakXT, Bruker, Germany). The flakes with a thickness of 110 nm and a lateral size of about 100 $\mu$m × 50 $\mu$m were chosen for further patterning. A negative tone electron beam resist (ARN 7520.18 from Allresist, Germany) was spin-coated on the sample at the speed of 4000 rpm and pre-baked on a 85$^o$C hotplate for 5 minutes (measured thickness 469 nm). A conductive polymer (AR-PC 5091.02 from Allresist) layer was successively spin-coated in order to prevent the surface charging during e-beam exposure. Then we used an EBL tool (Pioneer, Raith, Germany) for sample patterning. We used the 21 kV acceleration voltage of the electron gun, and the beam current was 0.24 nA for the aperture of 30 $\mu$m. The typical write field was 500 $\mu$m, and the nominal exposition dose was 1000 $\mu$C/cm$^2$. The pattern represented a square array of 50×50 dots with a period between dots of 2 or 4 $\mu$m. After the exposure, the conductive resist was rinced off in the pure deionized water, and the negative tone resist was developed in a AR 300-47 solution containing TMAH (tetramethylammonium hydroxide). The next step was to etch pillars into $MoS_2$ by RIE in $SF_6$ at 100 W power using the patterned photoresist as a mask. The etching rate of $MoS_2$ was evaluated as 100 nm/min by comparison with the reference nonpatterned $MoS_2$ flake of known thickness. The etching time was adjusted to completely etch $MoS_2$ between the pillars. Finally, the resist was removed in a corresponding organic solvent solution. Figure S1 shows SEM images of the flakes with the visible disk-like nanoresonators after fabrication.

**Experimental setup**

The nonlinear signal from the sample was detected using a home-built multiphoton microscope based on the femtosecond laser (Coherent Chameleon, 150 fs pulse duration, 80 MHz repetition rate) tunable in the spectral range from 680 nm to 1080 nm. Registration of the SH signal reflected from the sample was made in two ways. Firstly, the laser beam is focused at the sample surface and reflected SH signal is measured by a monochoromator and CCD camera (Andor Clara) to record the spectrum of the nonlinear response. Secondly, we use a kind of Köhler illumination scheme by installing an additional lens in the pump channel and observing a SH image of the sample from a large surface area on a CMOS camera (Photometrics Prime). The precise position of the sample relative to the pump spot is controlled by a 3-axis motorized stage. The modification of nonlinear setup also allows us to measure linear scattering and reflection spectra at fundamental wavelengths without manipulating the sample. The detailed description is given in Supplementary Information, Section II.

**Numerical simulation**

The numerical simulation of scattering cross-section and electromagnetic field distribution is performed by finite differences in time domain (FDTD) method in Lumerical software. The calculated system consists of a $MoS_2$ disk on a silicon substrate covered with a 290-nm-thick layer of silicon oxide. The in-plane and out-of-plane refractive indexes of $MoS_2$ are taken from Ref.[1]. The multipole mode decomposition of the scattering radiation is performed by numerical integration of the total electric field induced by a normally incident plane wave inside the nanoantenna [44]. The multipole mode decomposition for $MoS_2$ disk in free space is performed by discrete dipolar approximation (DDA) technique [45, 46].

**Table of Contents**

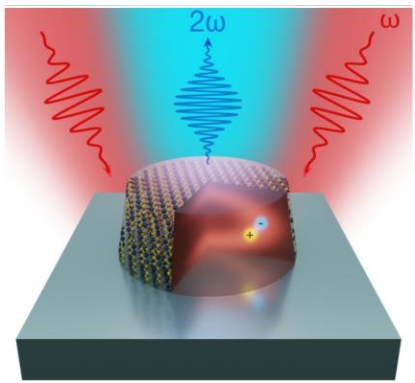

The tailoring of Mie resonances with excitons in $MoS_2$ nanodisk leads to 23-fold enhancement of second-harmonic generation intensity in respect to high-quality exfoliated $MoS_2$ monolayer. Second-harmonic signal demonstrates a strongly anisotropic response typical of a $MoS_2$ monolayer due to the single-crystal structure of the fabricated nanodisks, providing polarization degree of freedom to manipulate nonlinear response.